\documentclass{aa}  
\usepackage{txfonts}

\usepackage{natbib}
\usepackage{graphicx}
\usepackage{amsmath,amssymb}
\usepackage{ulem}
\usepackage{enumitem}

\newcommand{\ch}[1]{#1}

\newcommand{\be}{\begin{equation}}
\newcommand{\ee}{\end{equation}}

\begin{document}

\title{Alfv\'en-dynamo balance and magnetic excess in MHD turbulence}

\author{Roland Grappin
\inst{1}
\and
Wolf-Christian~M\"{u}ller
\inst{2}
\and
Andrea Verdini
\inst{3}
} 

\institute{LPP, Ecole Polytechnique, France\\
\email{Roland.Grappin@lpp.polytechnique.fr}
\and
Technische Universit\"at Berlin, Zentrum f\"ur Astronomie und Astrophysik, Germany\\
\and
Universit\`a di Firenze, Dipartimento di Fisica e Astronomia, Firenze, Italy and Royal Observatory of Belgium, SIDC/STCE, Brussels}

\date{\today}

\abstract
{3D Magnetohydrodynamic (MHD) turbulent flows with initially magnetic and kinetic energies at equipartition spontaneously develop a magnetic excess (or residual energy), as well in numerical simulations and in the solar wind. Closure equations obtained in 1983 describe the residual spectrum as being produced by a dynamo source proportional to the total energy spectrum, balanced by a linear Alfv\'en damping term. A good agreement was found in 2005 with incompressible simulations; however, recent solar wind measurements disagree with these results.
}
{The previous dynamo-Alfv\'en theory is generalized to a family of models, leading to simple relations between residual and total energy spectra. We want to assess these models in detail against MHD simulations and solar wind data.
}
{The family of models is tested against compressible decaying MHD simulations with low Mach number, low cross-helicity, zero mean magnetic field, without or with expansion terms (EBM or expanding box model).}
{A single dynamo-Alfv\'en model is found to describe correctly both solar wind scalings and compressible simulations without or with expansion. It is equivalent to the 1983-2005 closure equation but with critical balance of nonlinear turnover and linear Alfv\'en times, while the dynamo source term remains unchanged. The discrepancy with previous incompressible simulations is elucidated.
The model predicts a linear relation between the spectral slopes of total and residual energies 
$m_R = -1/2 + 3/2 m_T$. 
Examining the solar wind data as in \cite{2013ApJ...770..125C}, our relation is found to be valid whatever the cross-helicity, even better so at high cross-helicity, with the total energy slope varying from $1.7$ to $1.55$.}
{}

\maketitle

\section{Introduction}

3D Magnetohydrodynamic (MHD) turbulent flows with initial equipartition of magnetic ($E^M_k$) and kinetic ($E^V_k$) spectral energy density spontaneously develop a magnetic excess. In the Solar Wind, the magnetic excess in the $k^{-5/3}$ scaling range is largest in the cold, slow wind with no or small mean field (e.g., \cite{1991AnGeo...9..416G,Bruno:2007tw}). 
Recent measurements have revealed that when the relative cross-helicity $\sigma_c = 2\left< v \cdot b\right>/(\left<v^2+b^2\right>)$ is smaller than $0.6$, the residual energy $E^R_k=E^M_k-E^V_k$ adopts the scaling $k^{-2}$ \citep{2013ApJ...770..125C}, while the total energy $E^T_k=E^M_k+E^V_k$ scales as $k^{-5/3}$.

\ch{The origin of the magnetic excess in the solar wind has been attributed to different physical mechanisms: (i) remnants of quasi-stationary solar magnetic structures \citep{Bruno:2007tw}; (ii) formation and persistence of current sheets \citep{1986PhFl...29.2513M};
(iii) selective decay of ideal invariants \citep{1991PhFlB...3.1848S};
(iv) fully developed turbulence. In the latter case, which interests us here, the residual energy spectrum results from the competition between the linear damping of Alfv\'en waves by the local mean field, the Alfv\'en effect \citep{Kraichnan:1965p9279}, and} the magnetic stretching which is the source proportional to the total energy spectrum (\cite{1983A&A...126...51G} and \cite{Muller:2005p705}, later on MG05). We call it the Alfv\'en-dynamo scenario. 
\ch{The model allows (in its stationary version) to predict the residual energy spectrum, given the total energy spectrum. 
}
In particular, it predicts that the slopes of total energy ($m_T$) and residual energy ($m_R$) satisfy:
\be
m_{R} = -1 + 2 m_{T}
\label{restot}
\ee
which gives $m_R=2$ only if $m_T=3/2$, at variance with the solar wind case, which shows in average
$m_R \simeq 2$ and $m_T\simeq 5/3$.
\ch{It is important to remark that the model discussed here is not a cascade theory: the magnetic excess is not a inviscid invariant, it is assumed to be the passive by-product of the two effects mention above. So, the validity of our model is potentially more general than the validity of, say, the Kolmogorov regime (e.g., \cite{2010PhRvE..81a6318L}).}

Criticism has been addressed to the use of the edqnm approximation to derive the dynamo-Alfv\'en equation. Indeed, the small-scale dynamics in MHD turbulence should be dominated by motions perpendicular to the large-scale magnetic field, which in turn should strongly reduce the influence of the Alfv\'en effect which is a basic part of the theory (\cite{1993noma.book.....B}, end of Chapter 6). 
Several attempts have been made since then to include anisotropy into the edqnm closure \citep{2012PhPl...19j2310G,2012ASPC..459....3B}; however in the case of a strong cascade with no global mean field, they lead to the prediction $E^R_k=E^T_k$, at variance with our numerical findings and also the solar wind as we will see.

Our aim here is (i) to investigate whether one can recover the solar wind regime via numerical simulations of either standard \textit{compressible} 3D MHD equations or 3D EBM equations, that is, compressible MHD including expansion terms \citep{1993PhRvL..70.2190G,Grappin:1996ey};
(ii) find an appropriate framework for a Alfv\'en-dynamo scenario describing simulations and the solar wind that takes into account anisotropy.

We consider in this work the case with zero mean field and low cross-helicity, and small Mach number. Note that although the solar wind is rarely strictly in a zero mean field configuration ($b_{rms}/B_0 \simeq 0.5$ for a period range from 10 to 2 hours, $B_0$  being the mean magnetic field at the 10 hours scale, e.g., \cite{1989JGR....94.6899R}), the magnetic excess is dominant close to the heliospheric current sheet where the mean field is small, which is the configuration studied by \cite{2013ApJ...770..125C}.

To provide a diagnostic tool for our simulations, we first generalize formally the Alfv\'en dynamo balance derived from edqnm, by considering different simple expressions for the characteristic times of the dynamo source and the Alfv\'en sink.
These return different scaling predictions for the residual energy spectrum, given the spectral slope for total energy, quantitative predictions for the relative levels of both spectra, and also relaxation curves.
We test these different models against our simulations. 

We find that compressible MHD simulations, both standard and with expansion (EBM), show a quasi-stationary regime corresponding to that found in the solar wind. This quasi-stationary state is well described, qualitatively (scaling) and quantitatively (amplitude) if we keep the dynamo source term as in the 1983 edqnm scenario, but change the Alfv\'en sink, taking into account the critical balance between nonlinear coupling and linear propagation. The relation between the residual and total energy spectral slopes predicted by our theory applies not only to the zero cross-helicity wind, but also to high-speed streams with relative cross-helicity close to unity, and shallower spectral slopes.

We show that incompressible MG05 simulations differ from the compressible ones studied here insofar as the scaling law for the residual energy actually depends on the kind of spectrum used, whether reduced or isotropized. Such a dependence is not observed in the compressible simulations considered here. 


The plan is as follows. First we derive the family of Alfv\'en-dynamo models relating the residual and total energy spectra. 
Then we examine direct simulations of compressible MHD with low Mach number, both with and without expansion terms in the light of the different Alfv\'en-dynamo models. The last section is a discussion.

\section{Generalizing the Alfv\'en-dynamo model}
The Alfv\'en dynamo model is obtained starting from the incompressible MHD equations and using the edqnm spectral closure. It leads to a closed system of equations relating the different second-order moments of the system. The general form (with $E_k$ denoting either the kinetic, magnetic or residual energy spectrum) is $\partial E_k = \int dp dq k \theta E_p E_q$.
In the case of the residual energy, the integral may be separated in two, nonlocal and local, contributions, leading respectively to a linear damping term and a nonlinear source:
\be
\partial_t E^R_k = - \theta/{t_A^0}^2 E^R_k  + \theta/t_{NL}^2 E^T_k
\label{edqnm}
\ee
where the characteristic times $t_A^0$ and $t_{NL}$ are (the magnetic field is expressed in units of Alfv\'en velocity):
\be
t_A^0 = 1/(kb_0)
\label{iso}
\ee
\be
t_{NL} = 1/(k(u^2+b^2)^{1/2}) \simeq (k^{3/2} (E^{T}_k)^{1/2})^{-1}.
\label{tnl}
\ee
and $\theta$ is the relaxation time of triple correlations (in principle $\theta = t_A^0$, see below). 
\ch{Note that the form of the source term is local, while some authors (e.g., \cite{2005PhRvE..72d6301A}) claim that kinetic-magnetic exchange (other than the simple Alfv\'en effect) should show important non-local contributions. 
However, (i) \cite{Aluie:2010kv} have criticized the methodology of this work; 
(ii) a workable non-local model would be very difficult to build, and as we will see, it works well in an (albeit modified) local form.}

We generalize now the edqnm model by writing instead of eq.~\ref{edqnm}:
\be
\partial_t E^R_k = - E^R_k/t_{D} + E^{T}_k/t_{dyn}
\label{modgen}
\ee
where the two times scales, $t_D$ (damping time) and  $t_{dyn}$ (\textit{dynamo} time) can be chosen independently one from another. This leads to the equilibrium solution:
\be
E^R_k/E^T_k = t_{D}/t_{dyn}.
\ee
The edqnm model \ch{is described by eq.~\ref{edqnm} with $\theta=t_A^0$, and is thus equivalent to eq.~\ref{modgen} with} $t_{D}=t_A^0$ and $t_{dyn}=t_{NL}^2/t_A^0$, which leads to $E^R_k/E^T_k = (t_A^0/t_{NL})^2$.

Other expressions of the damping and dynamo times lead to the general family of residual-total energy relations:
\be
E^R_k/E^T_k = (t_A^0/t_{NL})^\alpha
\label{balance}
\ee
This leads to explicit relations between residual and total energy spectra after replacing the nonlinear time as a function of the total energy spectrum as in eq.~\ref{tnl}:
\be
E^R_k = b_0^{-\alpha} k^{\alpha/2} (E^{T}_k)^{1+\alpha/2}
\ee
Also, the following relation holds between spectral slopes ($E^{T}_k \propto k^{-m_{T}}, E^R_k \propto k^{-m_R}$):
\be
m_R = -\alpha/2 + m_{T}(1+\alpha/2)
\label{slopes}
\ee
Different values of $\alpha$ can be obtained by using the following expressions of the damping and dynamo timescales in terms of the basic timescales $t_A^0$ and $t_{NL}$:
\begin{align}
t_{D}&=t_{NL}  &t_{dyn} &= t_{NL} &\alpha=0
\label{a} \\
t_{D} &= t_A^0  &t_{dyn}&=t_{NL} &\alpha=1
\label{b} \\
t_{D}& = t_{NL}  &t_{dyn}&=t_{NL}^2/t_A^0&\alpha=1
\label{c} \\
t_{D}&= t_A^0  &t_{dyn} &= t_{NL}^2/t_A^0 &\alpha=2
\label{d}
\end{align}
As a rule (except perhaps at the largest scales), one has $t_A^0 < t_{NL} < t_{NL}^2/t_A^0$, with the inequality becoming stronger at small scales. 
The first scenario ($\alpha=0$, damping and dynamo both based on the nonlinear time $t_{NL}$) is the simplest of all \citep{2012PhPl...19j2310G}: it leads to an extreme regime with $E^R_k=E^T_k$ at all scales, far both from solar wind and numerical results as we will see.

The last scenario with fast damping and slow dynamo based on the long diffusive time  
$t_{dyn}=t_{NL}^2/t_A^0$, is the edqnm prediction studied in MG05; it leads to $\alpha=2$, hence to a fast decreasing residual spectrum with slope $m_R=7/3$ when $m_T=5/3$.

The intermediate scenarios (respectively fast damping and dynamo, and slow damping and dynamo) both lead to $\alpha=1$, and hence to $m_R=2$ when $m_T=5/3$: they are thus candidates for the description of the solar wind dynamics. 

\section{
Numerical results}
\label{results}
In this section we consider two decaying compressible simulations with zero mean field, low cross-helicity and resolution $512^3$: (i) 3D MHD (run A); (ii) 3D EBM (run B). In both cases, the initial Mach number is small (and remains so), thus minimizing compressible effects. Our main issue in this section is to assess eq.~\ref{balance}:
are the numerical results for equilibrium residual energy well described by one of the two algebraic models $\alpha=1$ or $\alpha=2$, and, accordingly, do the spectral scalings satisfy the corresponding eq.~\ref{slopes} ?

\subsection{Compressible turbulence}
The first run (A) has random isotropic initial conditions with $u_{rms}=b_{rms}=1$, $\nabla . u = 0$, equal magnetic and kinetic energies with spectra confined to $k \le 2$.
\ch{Initial relative cross-helicity is 0.17}, density is unity, pressure is uniform with sound speed  
$c_s=(5/3 \ P/\rho)^{1/2} \simeq 8$, initial Mach number $M_0 = u_{rms}/c_s \simeq 0.12$.
The domain is a cube of size $L_0 = 2 \pi$.

\begin{figure}[t]
\begin{center}
\includegraphics [width=\linewidth]{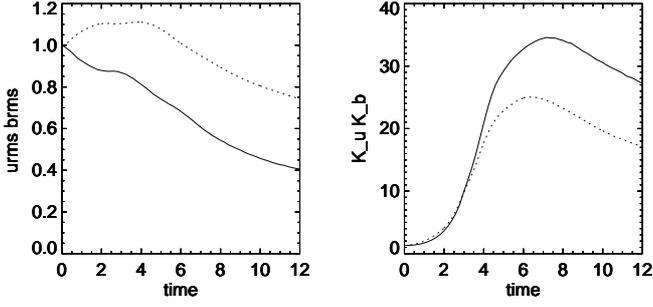}
\caption{
Run A (compressible MHD) - (a) Kinetic (solid line) and magnetic (dotted) rms amplitudes.
(b) Kinetic (solid line) and magnetic (dotted) Taylor wavenumbers vs time.
}
\label{fig1}
\end{center}
\end{figure}
The evolution of rms velocity (solid line) and magnetic (dotted line) fluctuations vs time is shown in the left panel of fig.~\ref{fig1},
during twelve nonlinear times $1/(k_0 u_{rms}(\text{t=0}))$.
The Taylor wavenumbers computed on the kinetic (solid) and magnetic (dotted) energy spectrum are shown in the right panel. 
One sees that the kinetic energy is transfered to magnetic energy in about two nonlinear times, while
small-scales wait up to about six nonlinear times to be fully excited, as indicated by the peaks of the Taylor wavenumbers. \ch{The relative cross-helicity (not shown in the figure) increases during the run from 0.17 to about 0.3}.

We consider now reduced 1D energy spectra $E(k_x)$, defined from the 3D spectral energy density $E_3$ as 
$E(k_x) = \int \int dk_y dk_z E_3(k_x,k_y,k_z)$
and the same for reduced spectra vs $k_y$ or $k_z$. Note that for run A with no expansion and no mean field, all directions should be equivalent; we thus use for all quantities the average of the three reduced spectra:
\be
E(k) = (1/3) (E(k_x)+E(k_y)+E(k_z))
\ee

\begin{figure}
\begin{center}
\includegraphics [width=\linewidth]{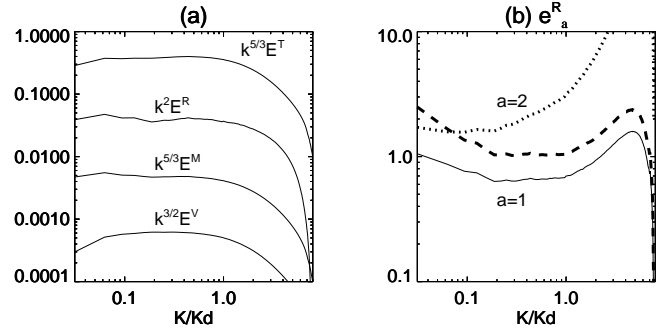}
\caption{Run A. Energy spectra averaged during time interval $10 \le t \le 12$. 
Wavenumbers are normalized by the dissipative wavenumber $k_d$ for which $k^2 E^M(k)$ is maximum, see text.
(a) From top to bottom: Total, residual, magnetic and kinetic energies, compensated resp. by $k^{-5/3}$, $k^{-2}$, $k^{-5/3}$, $k^{-3/2}$. Spectra are arbitrarily shifted vertically.
(b) Normalized residual energy $e^R_\alpha(k)$: 
(i) (solid) by the $\alpha=1$ model (eq.~\ref{balance});
(ii) (dotted) by the $\alpha=2$ model; 
(iii) (dashed) by the $\alpha=1$ model with the nonlinear time given by the advection time (see text)
}
\label{fig2}
\end{center}
\end{figure}

To reveal the spectral slopes of the different reduced spectra, we show in fig.~\ref{fig2}a the total, residual, magnetic and kinetic energies averaged during the time interval $10 \le t \le 12$, respectively compensated by the slopes $k^{-5/3}$, $k^{-2}$, $k^{-5/3}$, $k^{-3/2}$. The averages are made on 209 outputs, after expressing (by interpolating on a fixed grid) the spectra vs the ratio $k/k_d$, where $k_d$ is the wavenumber associated with a peak of the instantaneous spectrum of the current, namely of $k^2 E^M(k)$. Plateaux are seen to develop in the range $0.1 \le k/k_d \le 1$, thus showing good agreement with spectral slopes measured in the solar wind. The same set of slopes has been found recently in the 2D hybrid simulations by \cite{2015ApJ...804L..39F}.

Fig.~\ref{fig2}b allows to assess the Alfv\'en-dynamo scenario by showing the residual spectra normalized by the $\alpha-$model prediction with successively $\alpha=1$ and $\alpha=2$.
More precisely, one shows 
\be
e^R_\alpha(k) = E^R_k/(E^R_k)_\alpha
\label{normal}
\ee
where $(E^R_k)_\alpha=(t_A^0/t_{NL})^\alpha E^T_k$ is the equilibrium solution eq.~\ref{balance}. 

The solid curve shows the normalization by the $\alpha=1$ prediction, the dotted curve the normalization by the $\alpha=2$ prediction.
All spectra vs $k/k_d$ are again averaged over the 209 spectra stored during the time interval $10 \le t \le 12$. 
One sees that in the inertial range $0.1 \le k/k_d \le 1$, only the $\alpha=1$ normalization shows a plateau, indicating that the $\alpha=1$ scenario catches the basic physics, in agreement with the slopes
$(m_T,m_R)=(5/3, 2)$ obtained for the total and residual spectra in fig~\ref{fig2}a.

\begin{figure}
\begin{center}
\includegraphics [width=\linewidth]{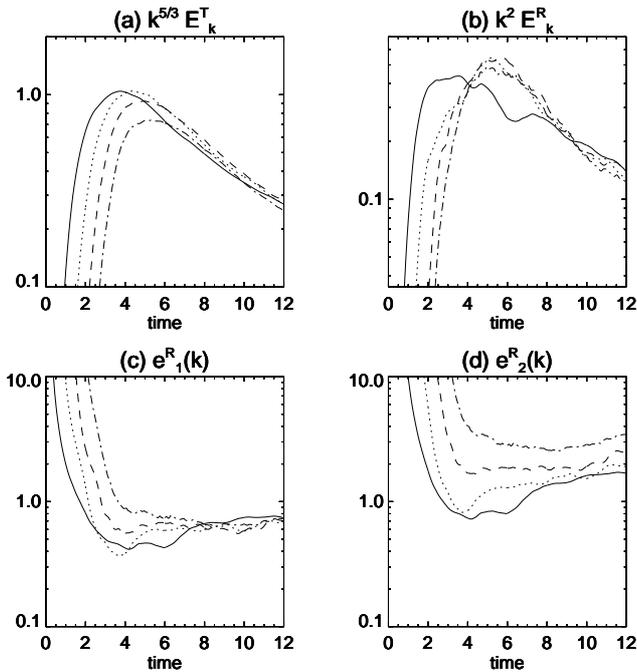}
\caption{Run A. Time variation of spectral energy density at specific wavenumbers: $k = 4, 8, 16, 32$, resp. solid, dotted, dashed, dotted-dashed lines.
(a) total energy times $k^{5/3}$ (b) residual energy times $k^{2}$ (c) residual energy $e^R_1(k)$ normalized by $\alpha=1$ equilibrium (d) residual energy $e^R_2(k)$ normalized by $\alpha=2$ equilibrium (see eq.~\ref{normal})
}
\label{fig3}
\end{center}
\end{figure}
A still better quantitative agreement can be obtained if
we use for the stretching time the advection time: $t_{NL}= 1/(ku)$ instead of eq.~\ref{tnl} (that allowed to close the problem). 
This gives a measured/predicted ratio closer to unity in the inertial range, as shown by the dashed curve in fig.~\ref{fig2}b.

To give an idea of how the spectra evolve with time, we show in fig.~\ref{fig3} the evolution of spectral energy density at five wavenumbers: 
$k = 4, 8, 16, 32$, for the four quantities: (a) total, (b) residual, (c) residual energy normalized by $\alpha=1$ model, (d) residual energy normalized by $\alpha=2$ model.
Total energy is multiplied by $k^{5/3}$, residual energy by $k^{2}$.
Total energies at $k=8, 16, 32$ are seen to collapse at about $t \ge 7$, revealing the formation of the $k^{-5/3}$ inertial range. At about the same time, the collapse of residual energy curves reveal the formation of the $k^{-2}$ range.
The modes 4 to 32, normalized by the $\alpha=1$ model, also show in panel (c) a nice collapse towards a value close to unity. 
In constrast, modes normalized by the $\alpha=2$ model remain significantly scattered.

\subsection{Turbulence with expansion}
We now consider with run B the 3D MHD equations modified by the expansion (EBM, \cite{1993PhRvL..70.2190G}).
\begin{figure}
\begin{center}
\includegraphics [width=\linewidth]{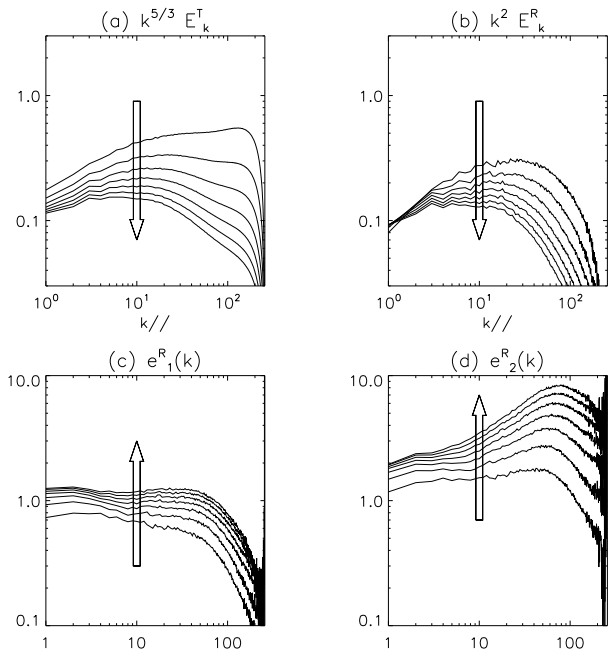}
\caption{Run B. Turbulence with expansion and  no mean field: spectra vs $k_R$ (radial wavenumber. i.e., in direction parallel to the mean radial flow) at $t = 0.8, 1.2, ... 3.2$ nonlinear times (vertical arrows indicate direction of time evolution).
(a) total energy spectra compensated by $k^{-5/3}$,
(b) residual energy, compensated by $k^{-2}$,
(c) residual energy $e^R_1(k)$ normalized by the $\alpha=1$ prediction, 
(d) residual energy $e^R_2(k)$ normalized by the $\alpha=2$ prediction (see eq.~\ref{normal}).
}
\label{fig4}
\end{center}
\end{figure}
The EBM equations allow us to follow the evolution of a plasma volume embedded in a uniform, given radial flow with speed U0 as in the solar wind. The radial flow cannot be eliminated by a plain Galilean transformation and forces the plasma volume to expand in directions perpendicular to the radial. 
This leads to many genuine effects specific of solar wind turbulence: (i) quadratic invariants are lost, and replaced by first order invariants (as mass, momentum, angular momentum, magnetic flux), (ii) cascade isotropy is lost, the turbulent cascade being stronger in the radial direction \citep{Dong:2014fi}.
\ch{Note that the expansion itself already provides a large-scale source of magnetic excess, as it selectively decreases only one component (the radial one) of the magnetic field in Alfv\'en speed units, and two components of the velocity field (e.g., \cite{1990JGR....9510291Z, 1995JGR...10014783O,
1993PhRvL..70.2190G, Dong:2014fi}).}

We consider initially an isotropic $1/k$ large-scale spectrum in order to mimic the large-scale fossil $1/f$ spectrum observed in the wind \citep{lrsp-2013-2}. The Mach number is $\simeq 0.1$ as for run A;
kinetic velocity and magnetic fluctuations are at equipartition, and $\nabla . u$=0.
The wind expansion rate normalized to the large scale nonlinear time is initially $\epsilon = \frac{U_0 L_0}{2\pi u_{rms}^0R_0}=2$, where $L_0$ is the initial size of the domain (with aspect ratio unity), and
$R_0$ the initial distance. 
This means that initially most of the spectral range (that with $k \ge 2$) has a nonlinear time shorter than the local transit time $t_e=R_0/U_0$. This leaves place for an inertial range to develop, together with specific spectral anisotropy characteristic of turbulence with expansion (see \cite{Dong:2014fi, Verdini:2015bx}).
Evolution is followed from time $t=0$ up to $t=3.2$ nonlinear times, corresponding to an heliocentric 
distance increase of $R/R_0=1+\epsilon t=7.4$ (also equal to the increase of the aspect ratio of the domain). 
\ch{During this time, relative cross-helicity remains smaller than $0.01$.}

As observational records by spacecrafts are made along the radial direction, leading via the Taylor hypothesis to 1D reduced spectra vs radial wavenumber, we present only such spectra in Fig.~\ref{fig4}, at times t=0.8, 1.2... 3.2. Total energy spectra are shown compensated by $k^{-5/3}$ in panel (a), residual spectra are shown compensated by $k^{-2}$ in panel (b).
On can thus see in panel (a) the progressive steepening of the total energy spectrum (starting from the initial $1/k$ spectrum) towards a $k^{-5/3}$ range and in panel (b) the formation of the $k^{-2}$ range for the residual spectrum.
Panels (c) and (d) show residual spectra normalized by respectively the $\alpha=1$ and $\alpha=2$ equilibrium spectra (eq.~\ref{balance}). 
The residual spectra are seen to converge towards the $\alpha=1$ model in the inertial range, much less so towards the $\alpha=2$ model.

Magnetic and kinetic spectra also exhibit (not shown) scalings close to respectively $5/3$ and $3/2$ but within a wavenumber range shorter than for previous run A.

\section{Discussion}
\subsection{The $\alpha=1$ model vs simulations/Solar wind}
Using both compressible MHD and EBM (i.e., compressible MHD modified by expansion) simulations with moderate Mach number, zero mean field, low cross-helicity, we have found that the $\alpha=1$ equilibrium between residual and total energy relation (eq.~\ref{balance}) holds, with the spectra showing specifically the slopes $(m_T, m_R)=(5/3,2)$.

We found that in run A the quasi-stationary magnetic spectrum scale as $k^{-5/3}$ and the kinetic spectrum as $k^{-3/2}$:  these scalings are observed in the solar wind \citep{Podesta:2007fu, Salem:2009dm}.
This is the first report of this set of slopes in simulations with low cross-correlation in 3D MHD simulations. Previously, the four spectral slopes had been found in 2D hybrid simulations by \cite{2015ApJ...804L..39F} with the same conditions (no mean field, weak compressibility, low cross-helicity). Also, incompressible 3D reduced MHD simulations have been reported with this set of slopes, but only when forcing with a large cross-helicity \citep{2011ApJ...741L..19B}. It is important to remark that the incompressible simulations are at variance with our compressible results: with no cross-helicity, these authors report $m^R=2$ together with $m^T=3/2$, which is the prediction of the $\alpha=2$ model, the one satisfied as well by the incompressible MG05 simulations.
This will discussed again below.
\begin{figure}
\begin{center}
\includegraphics [width=\linewidth]{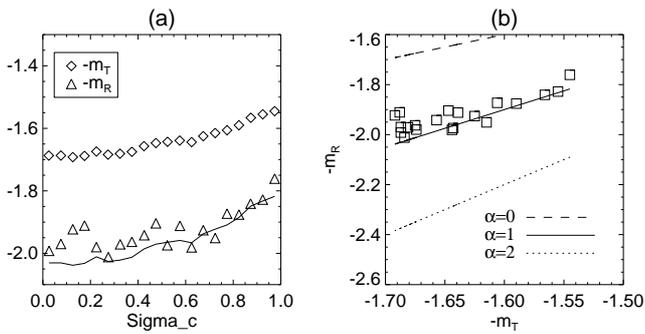}
\caption{Spectral slopes for total and residual energies: models vs observations. 
(a) Observations (symbols) as taken from fig5b from \cite{2013ApJ...770..125C} vs cross-helicity $\sigma_c$;
solid line: residual slope as predicted from observed total energy slope using eq.~\ref{magique} ($\alpha=1$ model)
(b) Scatter plot ($-m_T,-m_R$) from observations (squares);
straight lines: relation $m_R(m_T)$ given in eq.~\ref{slopes} with $\alpha=0$ (dashed line), $\alpha=1$ (solid), $\alpha=2$ (dotted).
}
\label{figchen}
\end{center}
\end{figure}

The scaling relation associated with the $\alpha=1$ model (eq.~\ref{slopes}):
\be
m_R = -1/2+(3/2)m_T
\label{magique}
\ee
applies well to the average slopes found in the solar wind: $m_T\simeq 1.7$, $m_R \simeq 2$ in the low cross-helicity wind. 
However, by examining the data published in \cite{2013ApJ...770..125C} (see their fig.~5b), we find that the agreement is actually more universal than that.
Eq.~\ref{magique}  works actually also for cross-helicities $\sigma_c$ larger than 0.6, for which the total energy spectrum becomes flatter.
This is seen in fig.~\ref{figchen}a, in which we reproduce the measured slopes for total and residual energies vs. $\sigma_c$ as in fig.5b of \cite{2013ApJ...770..125C}.
The solid line gives the predicted residual slope, replacing in eq.~\ref{magique} the total energy slope by its measured value. 
The agreement is seen to increase in the right part of the figure for large cross-helicity, which happens to be the region where error bars are the smallest (see original figure).

Fig.~\ref{figchen}b summarizes our findings by showing a scatter plot of measured slopes (residual slope vs total energy slope) and comparing with the predictions of the three models: $\alpha=0$ (dashed line), 1 (solid) and 2 (dotted). This allows to see how far from the true situation in the wind are, respectively,
the isotropic edqnm model ($\alpha=2$) on the one hand, and the simple $\alpha=0$ model.

This indicates that the Alfv\'en-dynamo theory is able to describe a large interval of turbulent parameters: it works as well in the balanced and imbalanced regime, with no mean field and with average mean field.

\subsection{Identifying separately source and damping}
To gain more insight into the dynamical process that leads to the $\alpha=1$ equilibrium, and in particular to discriminate between the two possible choices of characteristic times (eq.~\ref{b} or eq.~\ref{c}) that might rule the Alfv\'en-dynamo equation (eq.~\ref{modgen}), we now consider the response of the system without expansion (run A) to perturbations of the equilibrium residual energy in two successive experiments.
\begin{figure}
\begin{center}
\includegraphics [width=\linewidth]{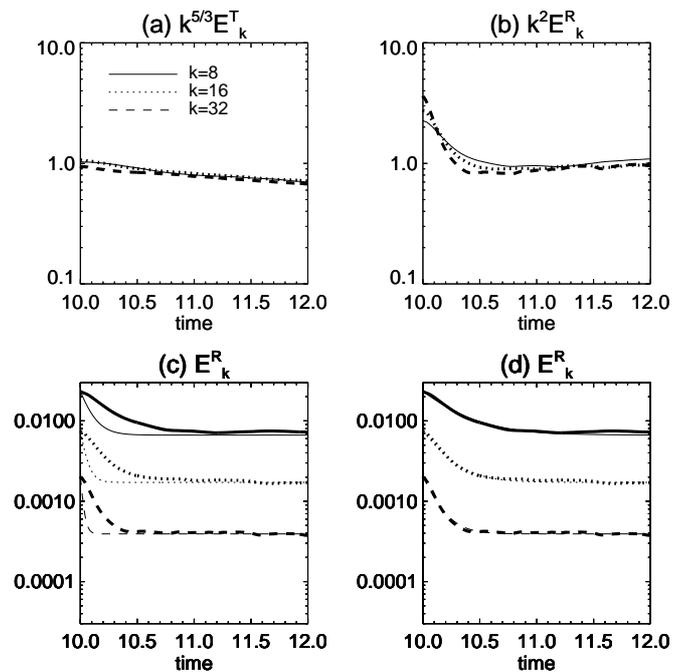}
\caption{Run $A_1$: return of residual energy to equilibrium after increasing the magnetic energy by a factor four in run A at time $t=10$. 1D reduced spectra, time evolution of modes $k=8, 16, 32$ (resp. solid, dotted and dashed lines):
(a) total energy, compensated by $k^{-5/3}$;
(b) residual energy normalized by the $\alpha=1$ solution evaluated at time $t=12$;
(c) residual energy, as well as the analytical solution with short damping time (thin lines, cf. eq.~\ref{solution} with $t_D=t_A^0$);
(d) same as (c) but with $t_D=t_{NL}$ for the analytical solution
}
\label{figa}
\end{center}
\end{figure}

In both experiments, we restart run $A$ at time $t=10$ after perturbing the residual energy spectrum, and follow the evolution up to $t=12$.
In the first experiment (run $A_1$), we increase the residual energy 
by increasing the magnetic energy by a factor four at all scales. 
In the second case (run $A_2$) we decrease the residual energy to zero by raising the kinetic energy to the level of magnetic energy in the whole spectral range.
\ch{Due to this procedure, the relative cross-helicity is raised to about 0.4, but stil remains below 0.45 up to the end of the run.}
\begin{figure}[ht]
\begin{center}
\includegraphics [width=\linewidth]{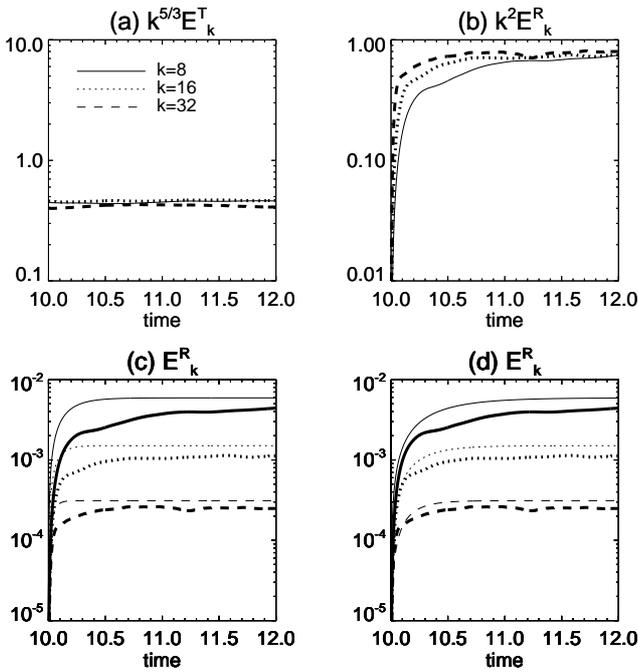}
\caption{Run $A_2$: return of residual energy to equilibrium after reducing the residual energy to zero by increasing the kinetic energy to the level of magnetic energy. Same caption as in previous figure.
}
\label{figb}
\end{center}
\end{figure}
The progressive relaxation of the system towards equilibrium is shown in the two figures~\ref{figa}-\ref{figb}. 
Panels (a) and (b) show respectively the behavior of modes $k=8, 16, 32$ for the total energy compensated by $k^{-5/3}$ and for the residual energy normalized by the $\alpha=1$ equilibrium.
One sees in panels (a) that the three modes of the total energy remain close together, indicating that the spectral scaling is basically not modified by the perturbation. Panels (b) show that the three modes of the normalized residual spectrum quickly recover the 
$\alpha=1$ solution in the time lapse $10-12$. This occurs  within a factor very close to unity in run $A_1$ (fig.~\ref{figa}), and within a factor $\simeq 0.8$ in run $A_2$ (fig.~\ref{figb}).

Last, we compare the measured numerical relaxation of the residual energy
to the analytical solution of the Alfv\'en-dynamo (eq.~\ref{modgen}) with $\alpha=1$.
To this aim, we rewrite eq.~\ref{modgen}, replacing the source term by the equilibrium solution of the $\alpha=1$ model:
\begin{align}
\partial_t E^R(k,t) &= (-E^R(k,t) +E^R_{eq}(k))/t_D
\end{align}
with $E^R_{eq}(k) = (t_A^0/t_{NL}) E^T(k)$,
and assume the total energy spectrum (as well as the caracteristic times $t_A^0$ and $t_{NL}$) 
to be time-independent, which is reasonably correct for runs $A_1$ and $A_2$.
In practice, we replace these parameters by their value at time $t=12$: $E^R_{eq}(k) = [(t_A^0/t_{NL}) E^T(k)]_{t=12}$.
The model solutions thus read
\be
E^R(k,t) = E^R(k,0) e^{-t/t_{D}} + E^R_{eq}(k) (1-e^{-t/t_{D}})
\label{solution}
\ee
with either $t_D=t_A^0$ (eq.~\ref{b}) or $t_D=t_{NL}$ (eq.~\ref{c}).

In panels (c) and (d) we show the model curves (thin lines) with respectively $t_D=t_A^0$ and $t_D=t_{NL}$, together with the numerical solutions.
The agreement for the three modes $k=8, 16, 32$ with the model $t_D=t_{NL}$ is clearly much better than with $t_D=t_A^0$.
It is perfect for run $A_1$ (fig.~\ref{figa}c), not as good for run $A_2$ (fig.~\ref{figb}), but still acceptable: the discrepancy is due to the fact that the new numerical equilibrium solution is a factor 0.8 smaller than the theoretical equilibrium solution (eq.~\ref{balance} with $\alpha=1$).

After having validated in Section~\ref{results} the ($\alpha=1$) equilibrium model, we have now shown that the relaxation time is $t_D = t_{NL}$. 
As a consequence, we deduce that the source term in eq.~\ref{modgen} reads
$E^T_k/t_{dyn}$ with $t_{dyn}=t_{NL}^2/t_A^0$
i.e., has the same form as in the edqnm solution.
In summary, we have proved that the magnetic excess results from the competition between two terms: (i) a \textit{slow} Alfv\'en damping with time scale equal to a nonlinear time as predicted by critical balance; (ii) a source term identical to that of the isotropic dynamo found in \cite{1983A&A...126...51G} and MG05. 

\subsection{Incompressible vs compressible simulations}
We finally come back on the MG05 incompressible simulations, which we have said to be well described by the $\alpha=2$ model, contrary to our compressible simulations that satisfy the $\alpha=1$ model.
\begin{figure}[ht]
\begin{center}
\includegraphics [width=\linewidth]{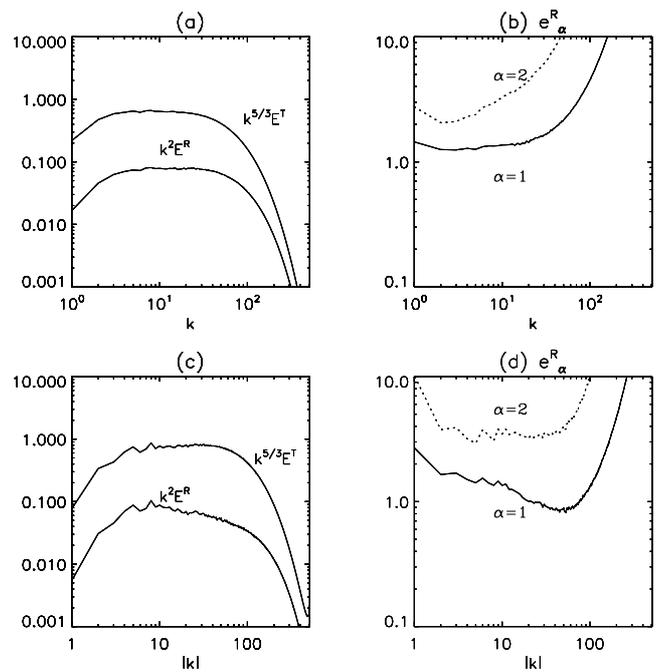}
\caption{Run C (incompressible MHD, no mean field, MG05). Spectra at time t=6.5.
Top: reduced spectra; Bottom: isotropized spectra.
Left: Total and residual energy spectra, compensated respectively by $k^{-5/3}$ and $k^{-2}$.
Right: Residual energy $e^R_\alpha(k)$ normalized by the $\alpha=1$ prediction (solid lines),
and normalized by the $\alpha=2$ prediction (dotted lines).
}
\label{fig5}
\end{center}
\end{figure}
We reexamined the MG05 simulations and found that the origin of the discrepancy lies in the method used to build 1D residual energy spectra. In MG05, the spectra $E^R(k)$ were built by averaging the spectral density within spherical shells (later on ``isotropized spectra'').
In contrast, in this paper, we used \textit{reduced} spectra, either $E^R(k_x)$ (run B) or averages of the three reduced spectra (run A).
While we expect that choosing reduced or isotropized spectra matters for run $B$ because expansion introduces a true physical anisotropy in the system (the radial direction becomes a symmetry axis), we expect nothing of the kind without expansion.
Indeed, we found that the choice of the spectrum (isotropized or reduced) makes no visible difference in scalings when dealing with runs $A$, $A_1$, $A_2$ presented here.

However, by reexamining the incompressible MG05 simulations, we found a different situation: \textit{residual} spectra (but not \textit{total} energy spectra) present different slopes, depending on whether they are reduced or isotropized.
Fig.~\ref{fig5} shows spectra for the incompressible run with no mean field considered in MG05 (denoted as run C). 
Top panels show reduced spectra, bottom panels show isotropized spectra, both at time t=6.5.

While the total energy spectra show $k^{-5/3}$ scalings whatever the spectrum used, the residual reduced and isotropized spectra show different scalings (compare panels (a) and (c)). 
The reduced residual spectrum scales as $k^{-2}$ (as do the compressible runs analyzed in the present work) while the isotropized residual spectrum scales as $k^{-7/3}$, as reported in MG05.
This is confirmed in panels (b) and (d) where we show the residual spectra normalized by the $\alpha=1$ (solid lines) and $\alpha=2$ (dotted lines) predictions (eq.~\ref{normal}). 
The $\alpha=1$ model indeed matches the reduced spectra while the $\alpha=2$ model matches the isotropized spectrum, at least qualitatively.
This clarifies the problem, but does not solve it: it remains that the incompressible simulations are singular with respect to the measurement of the residual energy spectrum. 

\subsection{Physical interpretation}
Let us come back on the list of different time scales in eqs.~(\ref{a}-\ref{d}) and forget for a moment that we know which model matches both our numerical (compressible) simulations and observational data.
Regarding the damping time $t_D$, the Alfv\'en effect is clearly at the origin of the damping of the magnetic excess. However, adopting the isotropized Alfv\'en time $t_D = t_A^0$ would contradict common wisdom that most of the energy lies in directions perpendicular to the local mean field \citep{1993noma.book.....B}. Indeed, for the majority of modes at a given scale $1/k$, the effective damping time is not equal to $t_A^0=1/(kb_0)$ but instead to $1/(k_\| b_0)$, which, by virtue of the so-called critical balance \citep{Goldreich:1995p4882}, is about equal to the nonlinear time $t_{NL}$:
this yields $t_D=t_{NL}$.
Now let us consider the dynamo time: the simplest choice is clearly $t_{dyn}=t_{NL}$ 
\citep{2012PhPl...19j2310G};
but since $t_D=t_{NL}$, this would imply that $E^R_k=E^{tot}_k$, in contradiction with solar wind observations and our numerical evidence.

In contrast, the remaining alternative, $t_{dyn} = t_{NL}^2/t_A^0$, (together with $t_D=t_{NL}$) matches our numerical data, and observations. 
\ch{Thus, the dynamo process responsible for the emergence and sustainment of the observed magnetic excess ( $\sim E^T/(t_{NL}^2/t_A^0)$) is   not straightforwardly connected with the energy cascade 
($\sim E^T/t_{NL}$), as it proceeds on a time scale much longer than the nonlinear time.}
This is not contradictory to what is known about the dynamo process, which also relies on a (long time) inverse cascade of magnetic helicity. However, this gives a prominent role to the isotropized Alfv\'en time, in contradiction 
with the critical balance between the effective Alfv\'en time and the nonlinear time. We have presently no solution for this paradox, as well as for the sensitivity, in incompressible solutions, of the residual spectrum to the definition of the spectrum (isotropized or reduced).

Other attempts to introduce anisotropy into the edqnm equation have made the \textit{a priori} assumption that the source of the residual spectrum has the same structure as that of the total energy spectrum \citep{2012ASPC..459....3B}. But, again, when applied to our case with no mean field, this assumption immediately leads to the invalid $\alpha=0$ prediction for which $m^R=m^T$, $E^R_k = E^T_k$. 

Several questions remain to be solved and are postponed to future work: (i) why incompressible simulations adopt a singular behavior, (ii) what is the origin of the long time scale $t_{dyn}$ of the local dynamo (eq.~\ref{c}).

\begin{acknowledgements}
We thank Simone Landi for interesting discussions. This work was performed using HPC resources from GENCI-IDRIS (grant 2015-040219). A. Verdini acknowledges partial funding from the Interuniversity
Attraction Poles Programme initiated by the Belgian Science Policy Office (IAP P7/08 CHARM) and from 
the European Union's Seventh Framework Programme for research, technological development and demonstration under grant agreement No. 284515 (SHOCK, http://project-shock.eu/home/).
\end{acknowledgements}

\bibliographystyle{aa}
\bibliography{grappin}

\end{document}